\documentclass{article}

\usepackage{arxiv}

\usepackage[utf8]{inputenc} 
\usepackage[T1]{fontenc}    
\usepackage{hyperref}       
\usepackage{url}            
\usepackage{booktabs}       
\usepackage{amsfonts}       
\usepackage{nicefrac}       
\usepackage{microtype}      
\usepackage{lipsum}

\usepackage{amsmath,amssymb,amsfonts}
\usepackage{graphicx}
\usepackage{textcomp}
\usepackage{xcolor}
\usepackage{subfig}
\usepackage{enumitem} 

\title{Synthesis of Realistic ECG using Generative Adversarial Networks}

\author{
  Anne Marie Delaney\thanks{Corresponding Author.} \\
  School of Computing\\
  Dublin City University\\
  Dublin, Ireland \\
  \texttt{anne.delaney32@mail.dcu.ie} \\
   \And
 Eoin Brophy \\
  Insight Centre for Data Analytics\\
  Dublin City University\\
  Dublin, Ireland \\
  \texttt{eoin.brophy7@mail.dcu.ie} \\
   \AND
   Tom{\'a}s E. Ward \\
   Insight Centre for Data Analytics\\
   Dublin City University\\
   Dublin, Ireland \\
   \texttt{tomas.ward@dcu.ie}\\
}

\begin{document}
\maketitle

\begin{abstract}
Access to medical data is highly restricted due to its sensitive nature, preventing communities from using this data for research or clinical training. Common methods of de-identification implemented to enable the sharing of data are sometimes inadequate to protect the individuals contained in the data.
For our research, we investigate the ability of generative adversarial networks (GANs) to produce realistic medical time series data which can be used without concerns over privacy. The aim is to generate synthetic ECG signals representative of normal ECG waveforms. GANs have been used successfully to generate good quality synthetic time series and have been shown to prevent re-identification of individual records.
In this work, a range of GAN architectures are developed to generate synthetic sine waves and synthetic ECG. Two evaluation metrics are then used to quantitatively assess how suitable the synthetic data is for real world applications such as clinical training and data analysis. Finally, we discuss the privacy concerns associated with sharing synthetic data produced by GANs and test their ability to withstand a simple membership inference attack.
For the first time we both quantitatively and qualitatively demonstrate that GAN architecture can successfully generate time series signals that are not only structurally similar to the training sets but also diverse in nature across generated samples. We also report on their ability to withstand a simple membership inference attack, protecting the privacy of the training set.

\end{abstract}


\section{Introduction}
Medical data is highly sensitive in nature. Because of this, access to and publishing of medical data is tightly controlled and regulated. Medical data required for secondary purposes such as software testing or clinical training must be anonymised. Most common methods for the de-identification of data are generalisation, randomisation or pseudonymisation \cite{Emam2015}. However, it has been shown that the de-identification of medical data does not guarantee the protection of privacy of all individuals in the data, and it is possible to re-identify individuals by linkage of data from other sources or from residual information \cite{ElEmam2011}. An alternative method of de-sensitising medical data is to generate synthetic data. It can be challenging to produce data that is representative of real medical data but protects the privacy of the individuals in the original data \cite{McLachlan2016,McSharry2003}. However, if successful, this data could be shared and published without privacy concerns for use in further research or clinical training.

Generative Adversarial Networks (GANs) have been used successfully to generate high quality synthetic images, with little work completed using GANs to generate synthetic time series \cite{Ledig2017,Reed2016,Radford2015}. A GAN consists of a generator and a discriminator. The generator \textit{G} is a neural network that takes random noise $\textbf{\textit{z}} \in \mathbb{R}^{r}$ and generates synthetic data. The discriminator \textit{D} is a neural network which determines if the generated data is real or fake. The generator aims to maximise the failure rate of the discriminator while the discriminator aims to minimise it. The GAN model converges when Nash equilibrium is reached. The two networks are locked in a two-player minimax game defined by the value function \textit{V(G,D)} (\ref{eq:1}), where \textit{D(\textbf{x})} is the probability that \textit{x} comes from the real data rather than the generated data \cite{Goodfellow2014}.

\setlength{\arraycolsep}{0.0em}
\begin{eqnarray}\label{eq:1}
\mathop{min}_{G} \mathop{max}_{D}V(G,D)&{}={}&E_{x \sim p_{data}(x)}[logD(\textbf{x})]+E_{z \sim p_{\textbf{z}}(z)}[log(1-D(G(\textbf{z})))]
\end{eqnarray}
\setlength{\arraycolsep}{5pt}

GANs aim to learn the underlying distribution of the training data. Thus, the synthetic data produced could be used as a substitute for sensitive data or to enrich smaller datasets. This is only possible if GANs prevent linkage of the training and synthetic data.

\begin{figure}[!htb]
    \center{\includegraphics[width=0.65\columnwidth,height = 4cm]
    {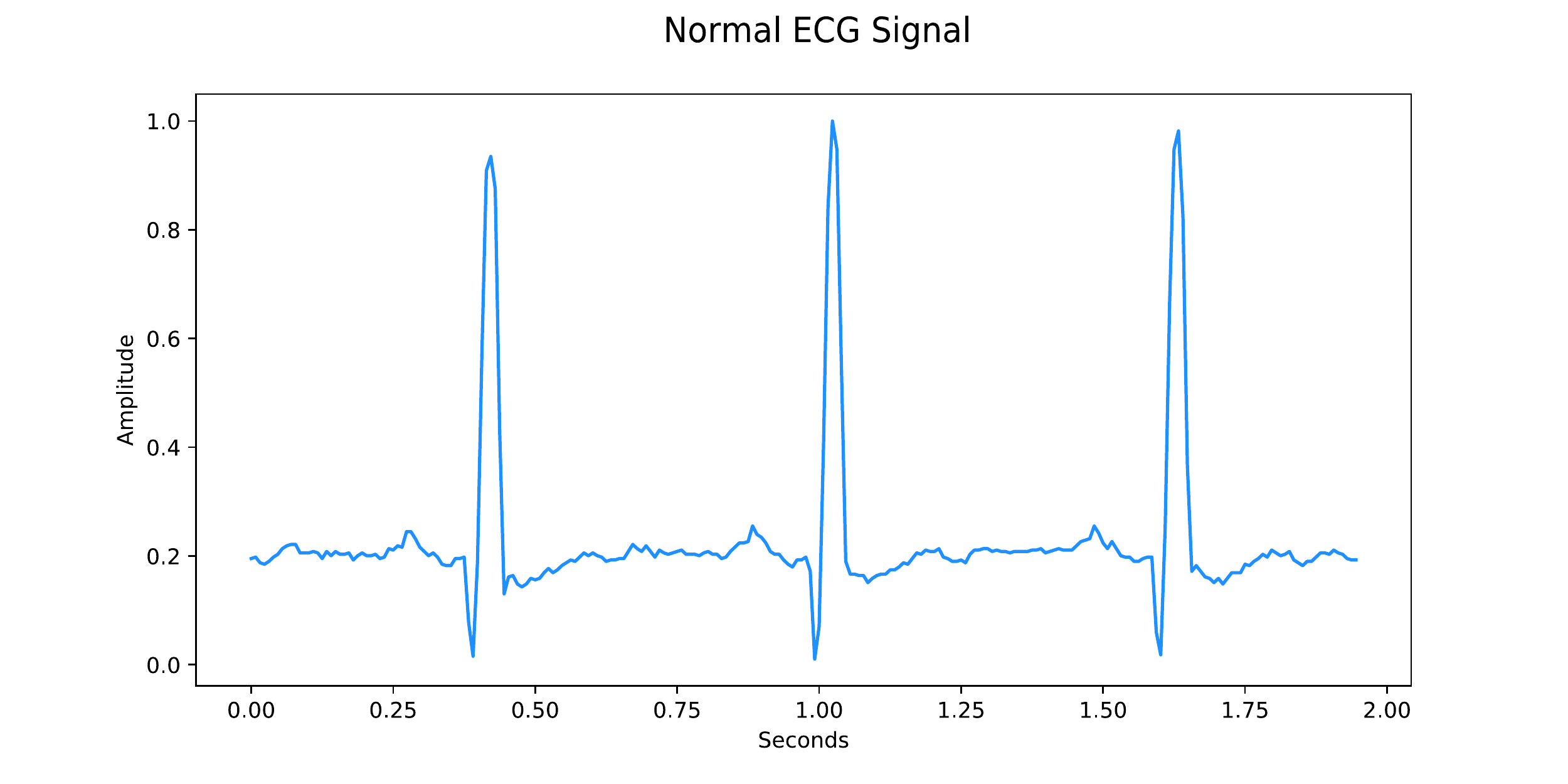}}
    \caption{\label{fig:normalECG}An example of normal ECG from the MIT-BIH Arrythmia Database.}
\end{figure}

The aim of our research is to use GANs to generate synthetic electrocardiogram (ECG) data representative of real ECG which could be made available for use in medical training or further research. An ECG is a recording of the heart's electrical activity during each cardiac cycle. A typical ECG signal can be seen in Figure \ref{fig:normalECG}. In this work, we focus on synthesising the lead II signal of normal ECG as found in the MIT-BIH Arrhythmia Database \cite{Moody2001, PhysioNet}.

Two largely-unsolved problems of GANs are their instability in training and the lack of suitable evaluation measures \cite{Creswell2018}. Many researchers rely on personnel with domain-specific knowledge to visually evaluate the synthetic data produced. This is both impractical and time-consuming and alternative methods are required. To assess the quality of the synthetic data generated in this study, both maximum mean discrepancy and dynamic time warping will be used. Their ability to determine the quality of the synthetic time series will be investigated to decide if they are suitable for GAN evaluation. Finally, a simple membership inference test will be used to assess if the data could be shared and published without privacy concerns.

The main objectives of our research are:
\begin{itemize}
\item Develop an algorithm which will generate real-valued time series data.
\item Investigate two evaluation metrics to assess the quality of the synthetic data produced.
\item Assess the privacy risk involved in using synthetic data generated by GANs as an alternative to the original data.
\end{itemize}

\section{Related Work}
It has been shown that common methods of de-identifying data do not prevent attackers from re-identifying individuals using additional data \cite{ElEmam2011,Malin2001}. Sensitive data is usually de-identified by removing the personal identifiable information (PII). However, work is on-going to create frameworks to link different sources of publicly available information together using alternative information to PII. Malin et al. (2001) developed a software program called REID to link individuals contained in publicly available hospital discharge data with their unique DNA records \cite{Malin2001}. Culnane et al. (2017) re-identified individuals in a de-identified open dataset of Australian medical billing records using unencrypted parts of the records and known information about individuals from other sources \cite{Culnane2017}. Hejblum et al. (2019) developed a probabilistic method to link de-identified Electronic Health Record (EHR) data of patients with rheumatoid arthritis \cite{Hejblum2019}. The re-identification of individuals in publicly available data sets can lead to the exposure of their sensitive health information. Health data has been categorised as special personal data by General Data Protection Regulation (GDPR) and is subject to a higher level of protection under the Data Protection Act 2018 (Section36(2)) \cite{GDPR36_2}. Consequently, it is important that research is carried out to find alternative methods of protecting sensitive health data to minimise the risk of re-identification.

Many different methods have been used in the past to generate synthetic data. In the medical domain, research has particularly focused on the generation of synthetic EHR \cite{McLachlan2016, Park2013}. Of particular relevance for our research are those methods which generate synthetic time series data. Previous approaches include the creation of dynamical models to produce synthetic ECG signals \cite{McSharry2003}. These models consist of three coupled ordinary differential equations with the user required to specify the characteristics of the heart rate signals to be generated. Many early methods require expert domain knowledge to generate synthetic data. More recent developments remove this dependency. An algorithm called WaveNet was developed which implemented an auto-regressive neural network that successfully generated synthetic music and speech \cite{VandenOord2016}. In other research, Dahmen and Cook (2019) developed SynSys to produce realistic home sensor data using hidden Markov models and regression models \cite{Dahmen2019}.

A significant development in the generation of synthetic data was the proposal of GANs. GANs are an alternative method of generating synthetic data which do not require input from domain experts. They were first proposed in the seminal paper by Goodfellow in 2014 where a multi-layer perceptron was used for both the discriminator and the generator \cite{Goodfellow2014}. Radford et al. (2015) subsequently developed the deep convolutional generative adversarial network (DCGAN) to generate synthetic images \cite{Radford2015}. They utilised convolutional neural network (CNN) architectures in their GAN and implemented techniques such as batch normalisation to improve stability in training. A recurrent GAN (RGAN) was first proposed in 2016. The generator contained a recurrent feedback loop which used both the input and hidden states at each time step to generate the final output \cite{Im2016}. Recurrent GANs often utilise Long Short-Term Memory neural networks (LSTMs) in their generative models to avoid the vanishing gradient problem associated with more traditional recurrent networks \cite{Hochreiter1997}. Since their inception in 2014, GANs have shown great success in generating high-quality synthetic images which are indistinguishable from the real images \cite{Guibas2017,Ledig2017,Reed2016}.

While the focus to date has been on the development of GANs for improved image generation, there has been a movement towards the use of GANs for time series and sequence generation. One such implementation involved the generation of polyphonic music as real-valued continuous sequential data using a LSTM in both the generator and discriminator \cite{Mogren2016}. In contrast, Yu et al. (2017) generated synthetic music by representing 88 distinct pitches with discrete tokens \cite{Yu2017}. This GAN, called SeqGAN, contained a LSTM in the generator with a CNN in the discriminator and out-performed alternative approaches for generating sequences of data. In the last year, a GAN producing raw-waveform audio (WaveGAN) was created by altering the architecture of the DCGAN \cite{Donahue2018}. It was used to generate realistic drum beats, bird song and  speech-like babble similar to those produced by unconditional autoregressive models. GANs were also used to generate single-channel electroencephalogram (EEG) data for motor movement in both the left and right hand \cite{Hartmann2018}. We are aware of only one work that implements both a GAN and a conditional GAN (CGAN) to generate real-valued multi-dimensional medical time series data \cite{Esteban2017}. A CGAN provides additional information to the generator and the discriminator to aid the creation of synthetic data \cite{Mirza2014}. Coinciding with the research discussed previously, Hyland et al. (2018) used a two-layer LSTM in both the discriminator and generator of the GAN and CGAN to create ECG data. The most recent attempt to generate synthetic ECG used bidirectional LSTMs in the generator and convolutional neural networks in the discriminator \cite{Zhu2019}. As part of our research, we have built on this work completed by Hyland et al. (2018) and Zhu et al. (2019).

A major challenge of GANs is that they are unstable to train and often result in mode collapse \cite{Creswell2018}. As the parameter space is high-dimensional and the loss function is non-convex, finding the Nash equilibrium is a challenge. Efforts have been made to identify ways to improve the stability of GAN training. Methods such as minibatch discrimination, batch normalization and one-sided label smoothing have shown some success in this area \cite{Salimans2016}. Alternative loss functions have been proposed such as the Wasserstein GAN which uses the Wasserstein distance to estimate the error of the discriminator in training \cite{Arjovsky2017}. Other research focuses on imposing constraints on the gradient of the discriminator \cite{Kodali2017}.

In addition, there is a lack of consensus in current research on the best choice of evaluation metric for GANs \cite{Creswell2018}. A wide range of evaluation metrics have been proposed to evaluate the quality of the data generated by GANs. This includes two-sample tests such as maximum mean discrepancy, alternative loss functions such as the Wasserstein distance and inception scores \cite{Borji2019,Im2018,Xu2018}. Hartmann et al. (2018) proposed that a combination of metrics should be used as each has advantages and disadvantages associated with their use \cite{Hartmann2018}. Hyland et al. (2018) implemented maximum mean discrepancy and two variations of the classifier two-sample test \cite{Esteban2017}. Zhu et al. (2019) implemented classical metrics associated with time series evaluation including the Frechet distance, root mean square error (RMSE) and percent root mean square method (PRD) \cite{Zhu2019}.

As well as evaluating the quality of the data, a wide range of methods have been used to evaluate the privacy risk associated with synthetic data created by GANs. Choi et al. (2017) performed tests for presence disclosure and attribute disclosure while others utilised a three-sample test on the training, test and synthetic data to identify if the synthetic data has overfit to the training data \cite{Choi2017} \cite{Esteban2017}. Work is on-going to develop machine learning methods with privacy-preserving mechanisms such as differential private GANs \cite{Abadi2016}. Xie et al. (2018) achieved differential private GANs by adding noise to the gradients of the optimizer during the training phase \cite{Xie2018}. Hyland et al. (2018) followed this approach and found that differential privacy training did not have an adverse effect on the quality of their generated ECG data \cite{Esteban2017}. Conversely, Hayes et al. (2018) carried out membership inference attacks on synthetic images and concluded that by improving the GAN's ability to minimise privacy risks, the quality of the data generated is sacrificed \cite{Hayes2019}.

\section{Model Design}
We propose a GAN based model to produce synthetic time series data. The models presented in this work follow the architecture of a regular GAN but utilise recurrent neural networks and convolutional neural networks which have shown previous success in synthesising time series data \cite{Esteban2017,Zhu2019}. Below we give a brief introduction to the architectures and evaluation metrics used in this study.

\subsection{Generative Adversarial Networks}  
For each GAN studied in this work, the input to the generator is a series of random noise $\textbf{z}_n \in \mathbb{R}^{T}$ of fixed length $T$ sampled independently from a normal distribution $\mathcal{N}(0,1)$. The generator then produces a series $G(\textbf{z}_n) \in \mathbb{R}^{T}$ of fixed length $T$. For each epoch during the training phase, the discriminator ingests real time series $\textbf{x}_n \in \mathbb{R}^{T}$ and fake time series $G(\textbf{z}_n)$ separately. The discriminator then classifies each series as real or fake and provides an output $D(\textbf{x}_n)$ and $D(G(\textbf{z}_n))$ for real and fake inputs respectively. In practice we feed batches of time series data to the generator and discriminator so that $Z \in \mathbb{R}^{m \times T}$ is a batch of $m$ series of random noise and $X \in \mathbb{R}^{m \times T}$ is a batch of $m$ real time series. For each series, the objective function $V(G,D)$ as in (\ref{eq:1}) is calculated and averaged over the batch. In practice, G is trained to maximise $log(D(G(\textbf{z}))$ instead of minimizing $log(1-D(G(\textbf{z})))$ as it provides stronger gradients early in training \cite{Goodfellow2014}.

\subsection{Recurrent Networks}
Recurrent Neural Networks (RNN) are neural networks that are used for series or sequence analysis. Long Short-Term Memory Networks are a type of recurrent neural network which solve the exploding or vanishing gradient problem associated with classical RNNs \cite{Hochreiter1997}. They consist of an input gate, forget gate and output gate. The cell state vector in the LSTM retains the relevant information from past steps in the sequence and is used in addition to the input and hidden state to generate the output. The operations performed in a LSTM cell are provided below where \(x_t\) is the input at time \(t\), \(h_t\) is the hidden state at time \(t\) and \(c_t\) is the cell state at time \(t\). 

\setlength{\arraycolsep}{0.0em}
\begin{eqnarray}
&f_t = \sigma_g(W_f x_t +U_f h_{t-1} +b_f)\\
&i_t = \sigma_g(W_i x_t + U_i h_{t-1} + b_i)\\
&o_t = \sigma_g(W_o x_t + U_o h_{t-1} + b_o)\\
&c_t = f_t \circ c_{t-1} + \sigma_c (W_c x_t + U_c h_{t-1} + b_c)\\
&h_t = o_t \circ \sigma_h(c_t)
\end{eqnarray}
\setlength{\arraycolsep}{5pt}

A bidirectional LSTM (BiLSTM) runs the input both forwards and backwards through time, thus preserving the context from both the future and the past \cite{Schuster1997}. In this work, the output of the forward and backward LSTMs is merged via summation at each time step.

\subsection{Convolutional Neural Networks}
Convolutional Neural Networks have shown great success in computer vision and natural language processing. In this work, one-dimensional CNNs are used as they are well-suited to time series data. CNNs consist of a convolution layer which applies a sliding filter over the time series in intervals defined by stride. The same filter is applied across all time steps which ensures that a feature is learned independently of its position in the series. This property of CNNs is known as weight sharing. The general form of the convolution is given by (\ref{eq:7})

\setlength{\arraycolsep}{0.0em}
\begin{eqnarray}\label{eq:7}
C_t = f(\omega \ast X_{t-\frac{l}{2}\colon t+\frac{l}{2}}  + b)       \forall t \in [1, T]
\end{eqnarray}
\setlength{\arraycolsep}{5pt}

where $C_{t}$ is the result of the convolution of an input \(X\) of length \(l\) with a filter \(\omega\), a bias parameter \(b\) and a non-linear activation function \(f\) at time $t$ \cite{Zhu2019}. By applying multiple filters, the network can learn multiple features of the data. The number of hidden neurons which fits the data is given by (\ref{eq:8})

\setlength{\arraycolsep}{0.0em}
\begin{eqnarray}\label{eq:8}
\frac{W-K+2P}{S}+1
\end{eqnarray}
\setlength{\arraycolsep}{5pt}

where W is the number of input dimensions, K is the kernel or filter size, S is the stride and P is the amount of zero-padding applied to the border of the data \cite{Zhu2019}. Next, a Rectified Linear Unit (ReLu) activation layer is applied to increase the non-linear properties of the features. Finally, a max pooling layer is utilised to reduce the spatial size of the representation. A filter is applied across the data via a sliding window, resulting in a feature space consisting of aggregations of the input data. The $j^{th}$ output of the max pooling layer with window size $a$ and stride $b$ is described by (\ref{eq:9}) below.

\setlength{\arraycolsep}{0.0em}
\begin{eqnarray}\label{eq:9}
p_{j} = max\{C_{bj+1-b},C_{bj+2-b},...,C_{bj+a-b}\}
\end{eqnarray}
\setlength{\arraycolsep}{5pt}

Multiple CNN layers can be used to capture the features of the data. CNNs are used in this study as they are often faster in training than LSTMs for longer sequence modelling as they do not have recurrent connections. When designing the convolution layers, small and odd-numbered filter sizes and strides were used as recommended in the literature \cite{Yamashita2018}. 

\subsection{Minibatch Discrimination}
A common result during the training phase of a GAN is mode collapse. This is when the generator learns a single solution which fools the discriminator. As the discriminator processes each point independently, it cannot know how dissimilar each solution of the generator is to the other. Thus, all outputs generated become the single point that the discriminator believes is realistic. One way of mitigating this behaviour is to include a minibatch discrimination layer in the discriminator. Minibatch discrimination examines solutions in combination and penalises the discriminator if the solutions are similar. It does this by transforming the processed input \(f(\textbf{x}_i) \in \mathbb{R}^{A}\) via a tensor \(T \in \mathbb{R}^{A\times B \times C}\) to create a matrix \(M_i \in \mathbb{R}^{B \times C}\). The output of the minibatch discrimination layer is defined as

\setlength{\arraycolsep}{0.0em}
\begin{eqnarray}\label{eq:10}
o(\textbf{x}_i)_b = \sum_{j = 1}^{n} [exp(-||M_{i,b}-M_{j,b}||_{L_1})] \in \mathbb{R}
\end{eqnarray}
\setlength{\arraycolsep}{5pt}

which is then concatenated with the input features \(f(\textbf{x}_i)\) \cite{Salimans2016}. This combination is fed into the next layer of the discriminator. In this way, the discriminator classifies each sample as real or fake, while using the auxiliary information describing how similar this sample is to the rest. In this work, the tensor $T$ is initialised with values sampled independently from the normal distribution $\mathcal{N}(0,1)$.

\subsection{Design of the Generator}
Two different generators were tested as part of this work. 
\begin{itemize}
    \item \textbf{LSTM:} This generator consists of two LSTM layers followed by a fully connected layer to generate the final output. There are 50 hidden units in both LSTM layers. A similar approach was used by Hyland et al. (2018) to generate synthetic sine waves, MNIST and aggregated health data \cite{Esteban2017}.
    \item \textbf{BiLSTM:} This generator consists of two BiLSTM layers with a fully connected layer to generate the final output. 50 hidden units were used in each BiLSTM. The final hidden state of the forward and backward LSTMs of the second BiLSTM layer are summed before ingested by the fully connected layer. Zhu et al. (2019) used BiLSTMs in their generator to generate ECG data \cite{Zhu2019}.
\end{itemize}

\subsection{Design of the Discriminator}
Two core architectures were investigated for use in the discriminator.
\begin{itemize}
    \item \textbf{LSTM:} This discriminator consists of two LSTM layers of 50 hidden units followed by a minibatch discrimination layer. Finally, a fully connected layer with a sigmoid activation function is used to for classification.
    \item \textbf{CNN:} This discriminator consists of a convolution-pooling layer with a ReLu activation function. This is followed by a minibatch discrimination layer, a fully connected layer and a sigmoid activation function for classification. Different numbers of convolution-ReLu-pooling layers were included as best suited the data.
\end{itemize}

\subsection{Evaluation metrics}
There is a lack of consensus as to how one should evaluate the data generated using GANs. For this research, a two-sample test called maximum mean discrepancy and a classical time series evaluation metric called dynamic time warping have been utilised to evaluate the GANs.

\begin{itemize}
\item \textbf{Maximum Mean Discrepancy:} Maximum mean discrepancy (MMD) evaluates the dissimilarity between two probability distributions $P_r$ and $P_g$ using samples drawn independently from each distribution \cite{Borji2019}. In practice we calculate the square of the MMD defined below.
    \begin{multline}
    MMD^{2} = \frac{1}{n(n-1)}\sum_{i= 1}^{n}\sum_{j \neq 1}^{n}K(x_i,x_j)-\frac{2}{nm}\sum_{i=1}^{n}\sum_{j=1}^{m}K(x_i,y_j)+\frac{1}{m(m-1)}\sum_{i=1}^{m}\sum_{j\neq1}^{m}K(y_i,y_j)
    \end{multline}
The larger the MMD statistic, the greater the dissimilarity between the distributions. We use the Gaussian RBF kernel $K(x,x') =  \sum_{j=1}^k e^{-\alpha_j \|x - x'\|^2}$ with bandwidth $\alpha$ equal to the pairwise distance between the joint data \cite{Sutherland2016}. The Python library torch-two-sample was used to calculate MMD.
    \item \textbf{Dynamic Time Warping:} Dynamic time warping (DTW) is a classical method for estimating the dissimilarity between two time series. It has been shown to be consistently superior to other similarity measures \cite{Serra2014}. DTW warps the the series along the temporal axis to optimally align the series and calculate the distance between them. The accumulated cost can be calculated by recursively applying
    \begin{eqnarray}
     D_{i,j} = f(x_i,y_j) + min\{D_{i,j-1},D_{i-1,j},D_{i-1,j-1}\}
    \end{eqnarray}
for \textit{i} in $1,...,N$ and \textit{j} in $1,...,M$ where \textit{N} and \textit{M} are the lengths of the series \textit{x} and \textit{y} respectively. Usually $f(x_i,y_j) = (x_i - y_i)^2$ \cite{Serra2014}. Due to the large volume of data studied in this work, FastDTW is used to approximate the DTW metric as it reduces the computational time required to calculate DTW to O(N) where N is the number of points in the series \cite{Salvador2007}.
\end{itemize}

Both evaluation measures are examined in this work to identify which most accurately describes the quality of the synthetic time series. 

\subsection{Privacy Evaluation}
Finally, a test for presence disclosure is included. This assesses the ability of the generator to generate data which protects the privacy of the training data. In this test, presence disclosure occurs if it is possible to determine that a particular record was used to train the GAN by observing the synthetic records. This is also known as membership inference attack \cite{Choi2017}. In this experiment, \textit{r} complete series are randomly sampled from both the training and test set. For each sampled series, its distance to all series in a synthetic data set is calculated. If a record in the synthetic dataset is within a defined threshold $\epsilon$ of the sampled record, they're claimed as a match. As the sample is drawn from the training and test set, the claimed match could be a true positive, (i.e. a particular record is correctly identified as a member of the GAN training set), a false positive (i.e. a particular record is incorrectly identified as a member of the GAN training set) or a true negative (i.e. we correctly claim that a particular record is not in the training set). The number of records sampled $r$ and the threshold $\epsilon$ are varied and the recall and precision calculated. Note that DTW could be used as the similarity measure to identify if a training record has merely been shifted or stretched by the generator to create synthetic data. However, due to the computational complexity of DTW, the Euclidean distance was used instead. DTW will be implemented for privacy evaluation as part of future work. 

\section{Method}
This section begins with a description of the computational platform and data used in this work. This is followed by a description of the GAN architectures studied and the approach used in the these experiments.\footnote[1]{Code for the 2CNN GAN available on Github: \url{https://github.com/Brophy-E/ECG_GAN_MBD.git}}

\subsection{Computing Platform}
An Nvidia Titan Xp GPU was used for the experiments in this work alongside Google Colaboratory. Colaboratory provides a GPU with 12-25GB of RAM for free use which was sufficient to train and validate all experiments using the sine wave and ECG data. The Python library Pytorch was utilised to build the neural networks as it provides strong support for GPUs and provides a user-friendly interface for working with tensors and Pytorch objects.

\subsection{Data}
To demonstrate the ability of GANs to generate realistic time series data, two different datasets were used. A comprehensive description of the data is given below.

\subsubsection{Sine Waves}
The aim of initial experiments was to generate sinusoidal waves. A simple time series was chosen for initial tests as they are easily evaluated via visual inspection. This provides a quick method of verifying that the GANs developed are capable of producing realistic time series. In addition to this, many biosignals such as EEG can be constructed as a sum of oscillatory components (sine waves). The training and test sine waves were generated with amplitudes in $[0.1,0.9]$, frequencies between $[2.0,6.0]$ radians and random phase between $[-\pi,\pi]$. The training set contains 10,000 waves while the test set contains 3000 sine waves. Both data sets consist of sequences of length 40.

\subsubsection{ECG Data}
The MIT-BIH Arrythmia Database was used for our experiments \cite{Moody2001, PhysioNet}. This database contains 46 half hour long ECG recordings of two-channel ambulatory ECG. Both normal ECG and a range of uncommon but significant ECG irregularities are included. Note that only the lead II signals for normal ECG were used in this study. 

A processed version of the MIT-BIH Arrythmia data is available online. The techniques listed below were applied to the original signals to clean the data in preparation for machine learning tasks \cite{Kachuee2018}. 

\begin{itemize}
\item The gain of 200 adu/mV was removed from the ECG series. 
\item Each patient's record was resampled from 360Hz to approximately 125Hz. 
\item The records were reshaped into series of 10 second duration.
\item Each series was normalised independently.
\item The R-peaks in each series were identified as those maxima which exceeded a threshold of 0.9.
\item The distance between each successive peak identified was calculated and the median $T$ determined.  
\item Series of length 187 were created by taking $1.2T$ successive datapoints beginning at each R-peak. This guaranteed at least one peak per series.
\end{itemize}

For our study, this dataset produced by Kachuee at al. (2018) was altered to create normal ECG series containing two R-peaks. Each normal series in the data set was concatenated with itself. The join was padded with the mean of the series to ensure that the series joined smoothly. The data was then resampled to approximately 60Hz to ensure the length of the new series was the same length as the original series. Subsequently, the training set consists of 72471 records while the test set contains 18118 records, each with series of length 187.

Initially, we generated a dataset containing series with two R-peaks using the raw data obtained directly from the Physionet Database. The data processing steps listed above were followed with small differences. For example, to create series containing two peaks, we used the median $T$ to create series of length $1.25T + 8$ beginning at $R_{i} - 8$ where $R_{i}$ is identified as the location of each R-peak. However, as there were no labels in our version of the database, we were unable to distinguish between the normal and abnormal ECG waves. Thus, the dataset produced by Kachuee at al. (2018) altered to include two R-peaks per series has been used in all experiments as it is possible to focus only on normal ECG.

\subsection{Generating Realistic Sine Data }
Combinations of the generator and discriminator architectures described in the Models section of this paper were used to generate synthetic sine data. The various GAN architectures studied in this work are given below.

\begin{itemize}
\item \textbf{LSTM GAN:} This GAN combines the LSTM generator and the LSTM discriminator.
\item \textbf{1CNN GAN:} This GAN combines the LSTM generator and a CNN discriminator with a single convolution-ReLu-pooling layer. 
\item \textbf{2CNN GAN:} This GAN combines the LSTM generator and a CNN discriminator with a pair of convolution-ReLu-pooling layers.
\item \textbf{1CNN BiLSTM GAN:} This GAN combines the BiLSTM generator and the CNN discriminator with a single convolution-ReLu-pooling layer.
\item \textbf{2CNN BiLSTM GAN:} This GAN combines the BiLSTM generator and the CNN discriminator with a pair of convolution-ReLu-pooling layers.
\end{itemize}

The parameters of the CNN layers are given in Table \ref{tab:CNNparameters1} where $C_i$ and $P_i$ refer to the $i^{th}$ convolution-pooling layer combination. Both 1CNN and 2CNN discriminators were investigated to determine if a second convolution-pooling layer improves the quality of the data generated.

\begin{table}[ht]
\begin{center}
\begin{tabular}{|l| c| c| c| c|c|} 
 \hline
 Layer & Input Size & Feature Maps & Filter &  Stride & Output Size \\
 \hline\hline
 C1 & 40    & 10 & 3 & 1 & 10*38\\ 
 P1 & 10*38 & 10 & 3 & 2 & 10*18\\
 C2 & 10*18 & 5 & 3 & 1 & 5*16\\
 P2 &  5*16 & 5 & 3 & 2 & 5*7\\  
 \hline
\end{tabular}
\caption{Parameters of each convolution-pooling layer in the CNN discriminator for sine wave generation.}
\label{tab:CNNparameters1}
\end{center}
\end{table}

Each GAN was trained for 120 epochs. For each epoch, the GAN was trained on the full training set in batches of 50 records. The discriminator was trained multiple times for a single generator training phase as recommended by Goodfellow et al. (2014) \cite{Goodfellow2014}. The Adam optimizer was used with learning rate $\alpha = 0.0002$ as it is computationally efficient and works well for deep learning \cite{Kingma2014}. Each GAN was first trained without a minibatch discrimination layer. Next, a minibatch discrimination layer was added to the discriminator with number of outputs set to 3, 5, 8 and 10 to determine how the number of outputs affects the data generated.

To assess the quality of the data generated, the MMD was calculated between each record in the test set and a synthetic data set of equal size. Average DTW was determined using a random sample of 13\% of both the test and synthetic data due to the large computational time required to calculate DTW. MMD and DTW were calculated at the end of each training epoch. To assess presence disclosure, the sample size $r$ varied between [250,3000] records while the threshold $\epsilon$ ranged from [0.05,0.5] of the mean Euclidean distance between all samples. A synthetic dataset of equal size to the training set was used for this test.

\subsection{Generating Realistic ECG Data }
Two GAN architectures were used to generate ECG data. They are described below.
\begin{itemize}
\item \textbf{4CNN GAN:} This GAN combines the LSTM generator and a CNN discriminator with four convolution-ReLu-pooling layers. 
\item \textbf{4CNN BiLSTM GAN:} This GAN combines the BiLSTM generator and the CNN discriminator with four convolution-ReLu-pooling layers.
\end{itemize}

The parameters of all four CNN layers are given in Table \ref{tab:CNNparameters2} where $C_i$ and $P_i$ refer to the $i^{th}$ convolution-pooling layer combination. Four CNN layers were used as the ECG data has greater complexity which could not be captured with one or two convolution-pooling layers. Each GAN was trained for 60 epochs. For each epoch, the GAN was trained on the full training set in batches of 119 records. As before, the discriminator was trained multiple times for a single generator training phase and the Adam optimizer was used. Each GAN was also trained with and without the minibatch discrimination layer and the number of minibatch discrimination outputs was varied as before.

Once again, due to the limited computational power available, it was not possible to calculate MMD or DTW with the complete test set. MMD was calculated using a random sample of 65\% of both the test and synthetic datasets, while the average DTW was calculated using 13\% of the test and synthetic datasets. MMD and DTW were calculated at the end of each training epoch. Finally, to test for presence disclosure, the sample size $r$ was varied between [1000,10000] records while the threshold $\epsilon$ ranged from [0.05,0.5] of the mean Euclidean distance between all samples. A synthetic dataset of equal size to the training set was used.

\begin{table}[ht]
\begin{center}
\begin{tabular}{|l| c| c| c| c|c|} 
 \hline
 Layer & Input Size & Feature Maps & Filter &  Stride & Output Size \\
 \hline\hline
 C1 & 187    & 3 & 3 & 1 & 3*185\\ 
 P1 & 3*185 & 3 & 3 & 1 & 3*185\\
 C2 & 3*185 & 5 & 3 & 1 & 5*181\\
 P2 &  5*181 & 5 & 3 & 2 & 5*90\\  
 C3 & 5*90 & 8 & 3 & 2 & 8*44\\ 
 P3 & 10*44 & 8 & 3 & 2 & 8*21\\
 C4 & 10*21 & 12 & 5 & 2 & 12*8\\
 P4 &  5*8 & 12 & 5 & 2 & 12*2\\  
 \hline
\end{tabular}
\caption{Parameters all four convolution-pooling pairs in the 4CNN discriminator.}
\label{tab:CNNparameters2}
\end{center}
\end{table}

\section{Results}
\subsection{Sine Wave Generation}
In this section, the results of all efforts to generate synthetic sine waves are discussed. For each GAN architecture listed above, the error plot, MMD and DTW were investigated to assess the quality of the data generated. Finally, the results of the presence disclosure test are also included.

\subsubsection{Error Plot}
\begin{figure}[!htb]
\centering
\subfloat[]{\includegraphics[width=0.75\columnwidth, height = 4cm]{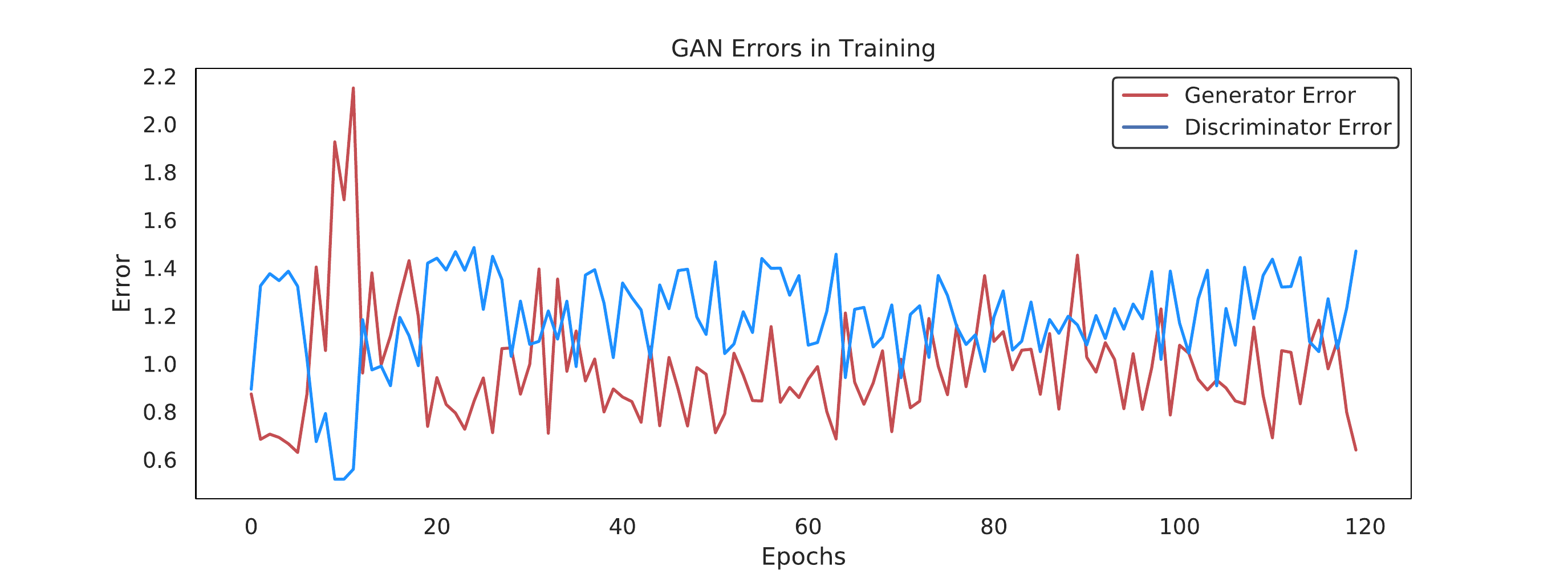}}\\
\subfloat[]{\includegraphics[width=0.75\columnwidth, height = 4cm]{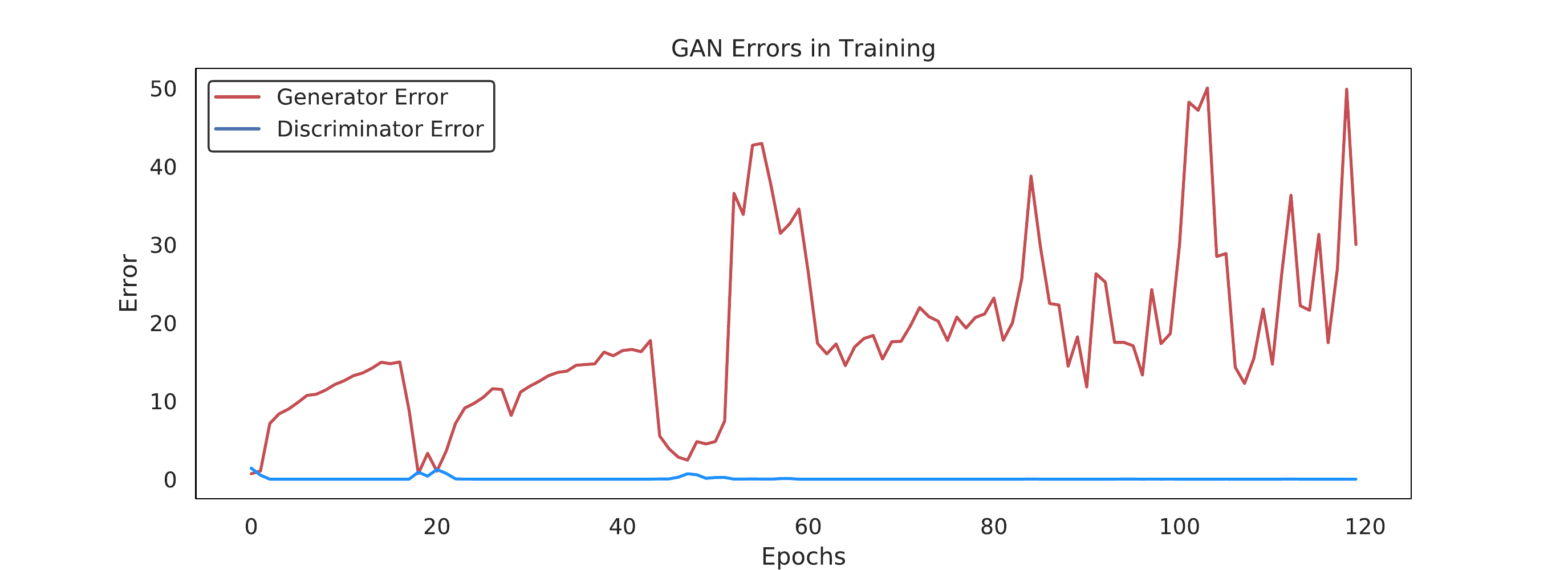}}
\caption{\label{fig:Errorplots}Examples of GAN error plots for (a) successful training for 1CNN BiLSTM GAN and (b) failed training for LSTM GAN.}
\end{figure}

In an optimum solution, both the generator loss and the discriminator loss converge to 1, where the discriminator loss is the sum of the average loss calculated independently for real and generated series. However, in practice this is almost impossible to achieve. For each experiment in this work, the generator and discriminator loss for each epoch was plotted to identify if the training phase was successful. Upon investigation, it was clear that convergence in the error plot is not indicative of high-quality synthetic data. However, the error plot was used as a means of identifying if training failed. If the losses explode as training progresses, this indicates that the GAN has failed to learn during the training phase. An example of the error plot for a failed GAN can be seen in Figure \ref{fig:Errorplots} (b).

\subsubsection{Evaluation of Generated Data}
\begin{figure}[!htb]
    \center{\includegraphics[width=0.75\columnwidth]
    {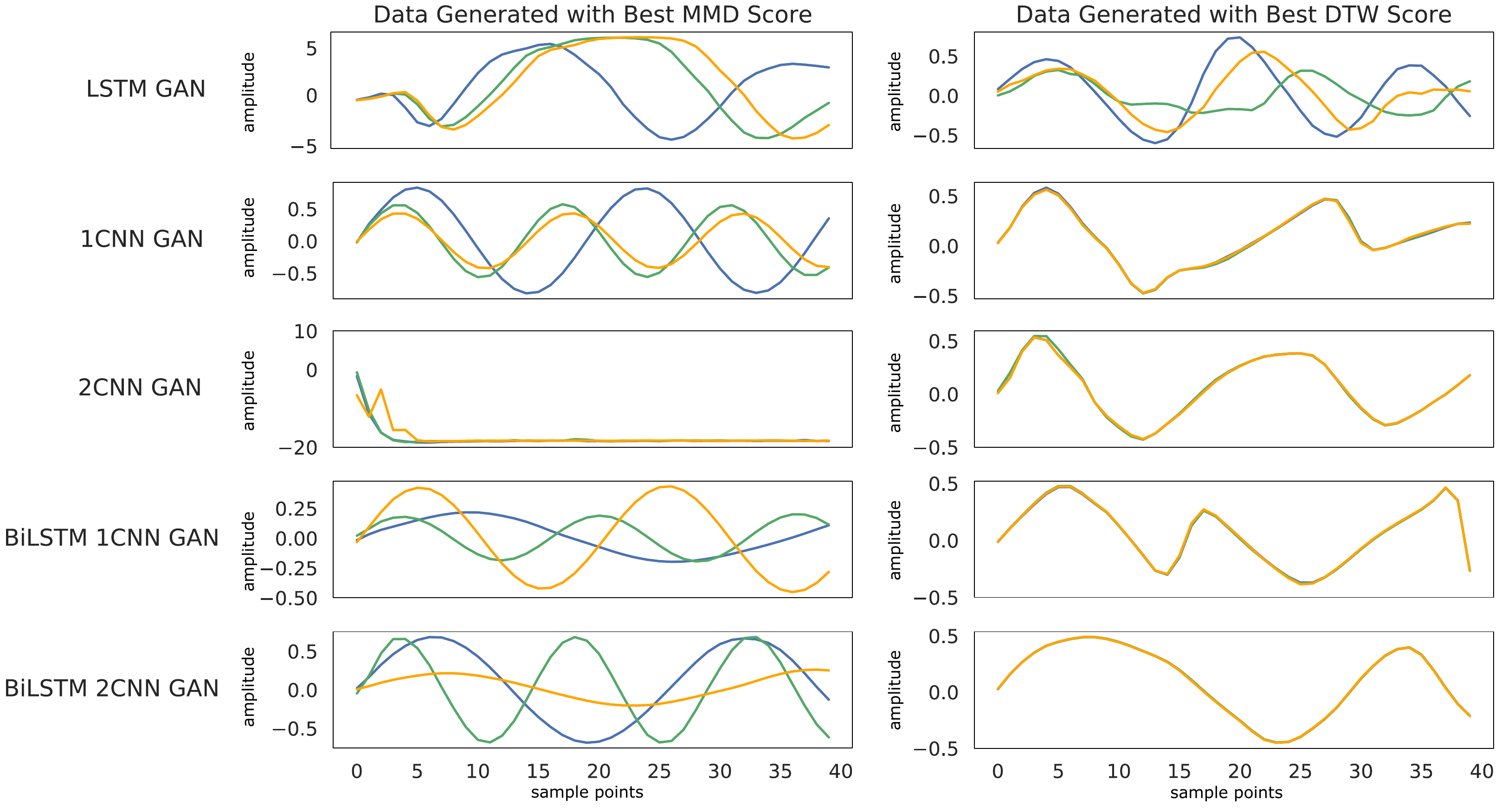}}
\caption{\label{fig:SineResults}Examples of sine waves corresponding to the MMD and DTW for each GAN reported in Table \ref{tab:SineMetrics} are displayed.}
\end{figure}

The best results for MMD and DTW obtained for each GAN architecture investigated are listed in Table \ref{tab:SineMetrics} and the corresponding series produced are displayed in Figure \ref{fig:SineResults}. The number of outputs in the minibatch discrimination layer used to produce these results is also included.

\begin{table}[ht]
\begin{center}
\begin{tabular}{|c| c| c|} 
 \hline
 GAN &  MMD (mb)& DTW  (mb)\\
 \hline\hline
 LSTM GAN & $3.43 \times 10^{-3} (10)$ & $7.998 (3)$\\ 

 1CNN GAN & $1.22\times 10^{-3} (10)$ & $7.497 (0)$ \\

 2CNN GAN & $\mathit{0.67\times 10^{-3} (8)}$ & $7.502 (0)$\\
 
 1CNN BiLSTM GAN & $\mathbf{1.05\times 10^{-3} (3)}$& $\mathbf{7.296 (0)}$ \\
 
 2CNN BiLSTM GAN & $2.63\times 10^{-3} (5)$ & $7.598 (0)$ \\
 \hline
\end{tabular}
\caption{\label{tab:SineMetrics}The best MMD and DTW obtained for each GAN is listed. The number of outputs in the minibatch discrimination layer is in brackets. Zero means no minibatch discrimination layer was included. The best MMD and DTW overall are in bold. The italicised results are generated series which obtained a good score but are not sine waves.}
\end{center}
\end{table}

In general, MMD was found to be a better assessment of the quality of data than DTW. It is possible that DTW behaves badly due to the oscillatory nature of sine waves. DTW attempts to warp the data along the temporal axis and this could be hampered by the periodic pattern. MMD tends to be lower when a diverse range of high-quality results are produced by the generator (i.e. minibatch discrimination layer is included), while DTW favours those GANs which produce many of the same result (are in mode collapse). However, MMD is insensitive to the range of the data and fails in certain cases, producing extremely low MMD ($< 0.001$). For example, it is clear in Figure \ref{fig:SineResults} that while the data produced by the 2CNN GAN has a very low MMD, they are not sine waves and have a much larger amplitude when compared to the original data. In comparison, DTW preserves the range of the data as it measures the distance between the series.

\begin{figure}[!htb]
    \center{\includegraphics[width=0.65\columnwidth]
    {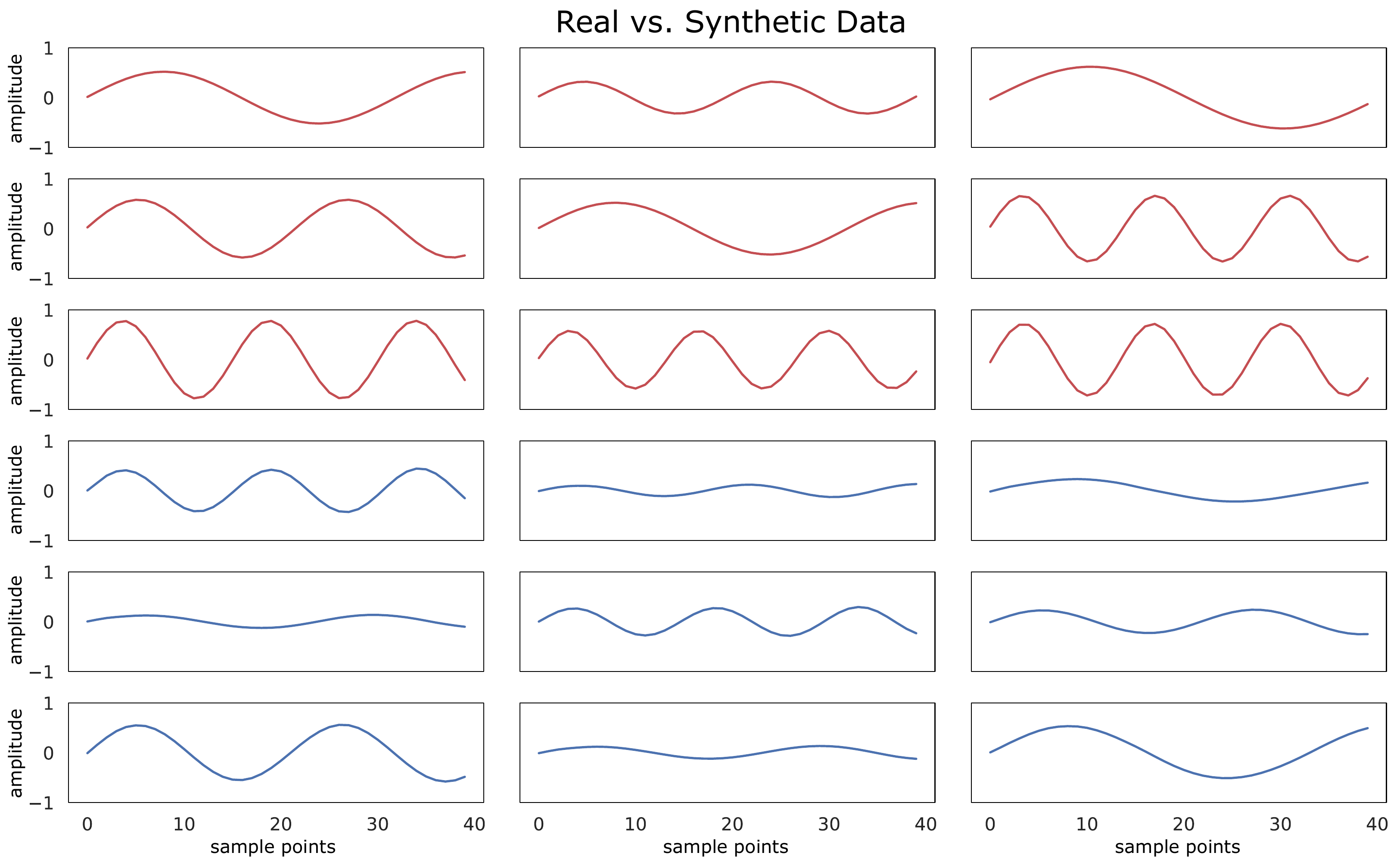}}
    \caption{\label{fig:real&fake}Examples of both real (red) and synthetic (blue) sine waves generated by the best run of the 1CNN BiLSTM GAN.}
\end{figure}

Using both the MMD statistic and visual inspection of the generated data, we can conclude that the 1CNN BiLSTM GAN produced the best results. In Figure \ref{fig:real&fake}, the real sine waves are indistinguishable from the synthetic waves produced by this GAN. This also shows that a single convolution-pooling layer is sufficient to learn the sinusoidal behaviour. In practice, the GANs using the 2CNN discriminator were more unstable in training than the 1CNN counterpart and did not improve the quality of the data generated. Finally, the LSTM GAN performed badly and thus was not used to generate ECG data in later experiments.

\subsubsection{Minibatch Discrimination}
\begin{figure}[!htb]
    \center{\includegraphics[width=0.75\columnwidth]
    {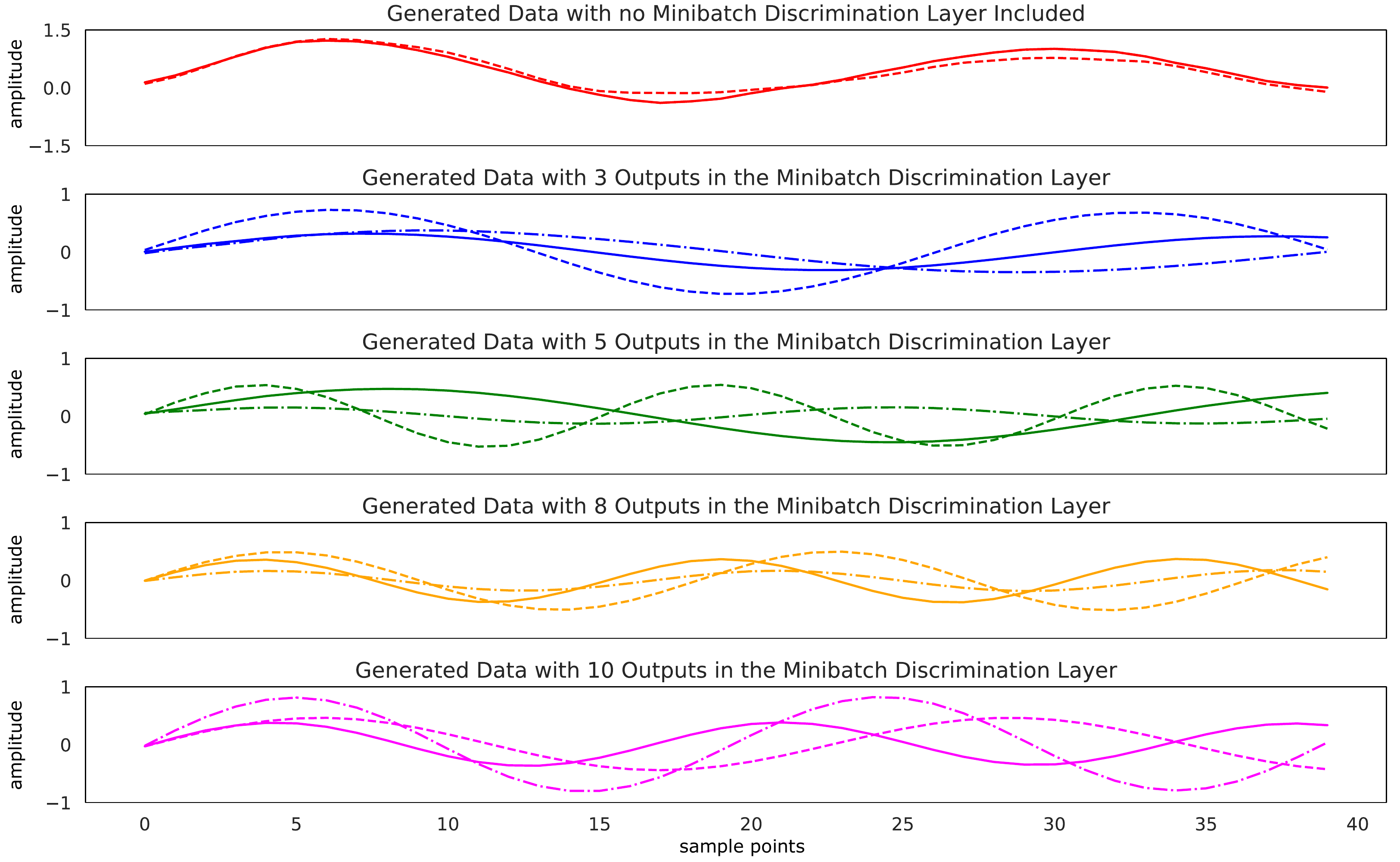}}
    \caption{\label{fig:mbinvestigation}Examples of sine waves produced by 1CNN GAN with increasing number of outputs in the minibatch discrimination layer. The diversity of waves produced is greater when a minibatch discrimination layer is included.}
\end{figure}

Minibatch discrimination was introduced in all experiments to prevent mode collapse and ensure diverse outputs were produced. It was found that in the absence of minibatch discrimination, the output of the GANs tended to be near or exactly identical. However, when minibatch discrimination was included, a wider diversity of sine waves were generated. This can be seen in Figure \ref{fig:mbinvestigation}. The inclusion of a minibatch discrimination layer also resulted in lower MMD scores as is clear in Table \ref{tab:SineMetrics}. However, there was no discernible improvement in quality with the inclusion of a higher number of outputs in the minibatch discrimination layer. 

It is worth noting that the inclusion of a minibatch discrimination layer increased the instability of the training phase for the LSTM GAN only. This could be a result of the concatenation of recurrent outputs of the LSTM with the static outputs of the minibatch discrimination layer. As the number of minibatch discrimination layer outputs was increased, the instability of the training phase also increased until the LSTM GAN with 10 minibatch discrimination outputs never converged.

\subsubsection{Presence Disclosure}
The data generated by the 1CNN BiLSTM GAN was tested to determine if this GAN is successful in protecting the privacy of the underlying training data. 
\begin{figure}[!ht]
\centering
{\includegraphics[width=0.75\columnwidth]{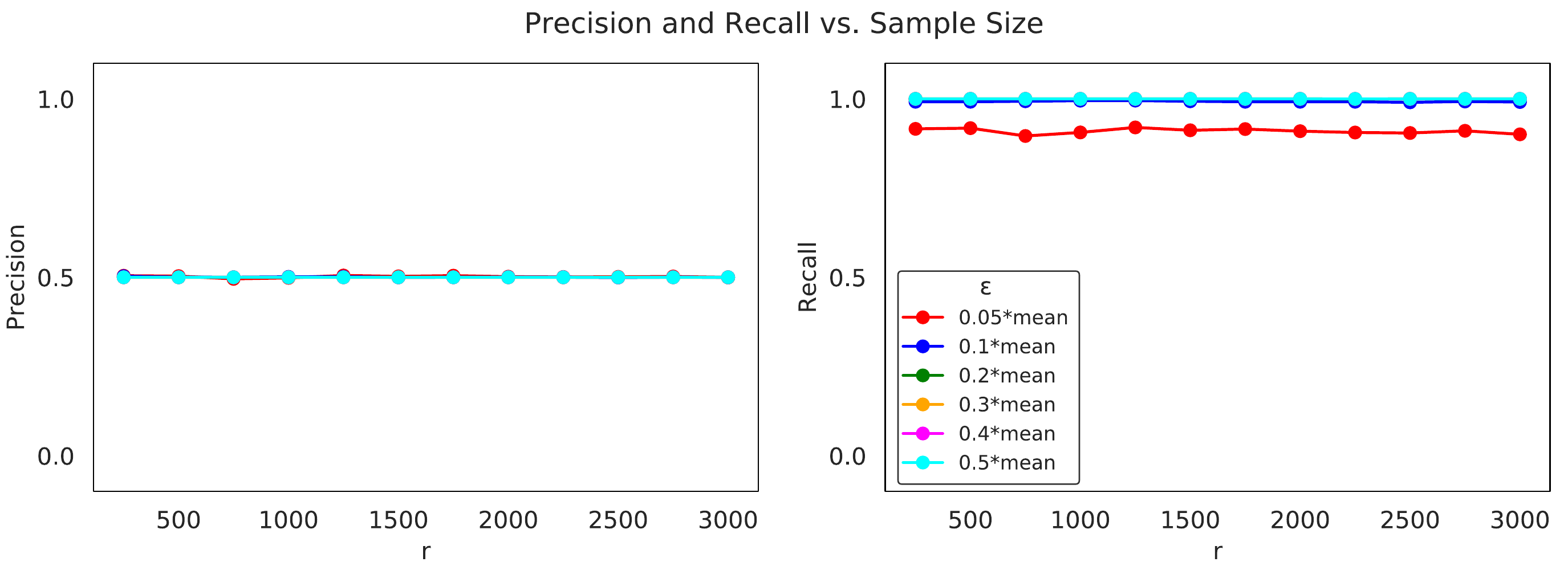}}
\caption{The precision and recall for the presence disclosure test using sine wave data.}
\label{fig4:SinePresence}
\end{figure}

In this case, x\% recall means that of those records that the attacker knows, the attacker managed to correctly identify that x\% of them were in the training set. x\% precision can be interpreted as of those that the attacker identified as being in the training set, only x\% of them were actually in the training set. As recall is almost 100\% for all distance thresholds $\epsilon$ and increasing sample sizes $r$, any leaked training records will be identified. However, as precision is 50\%, this information is almost useless as they will have also incorrectly identified an equal number of records as having been included in the training set. The high recall score may be due to the periodic nature of the sine waves increasing the possibility that two waves are identified as matches.

\subsection{ECG Generation}
In this section, the results of all experiments attempting to generate synthetic ECG waves are discussed. As before, for each GAN architecture investigated, MMD and DTW are reported and the presence disclosure test results included.
\subsubsection{Evaluation of Generated Data}
\begin{figure}[!ht]
    \center{\includegraphics[width=0.75\columnwidth]
    {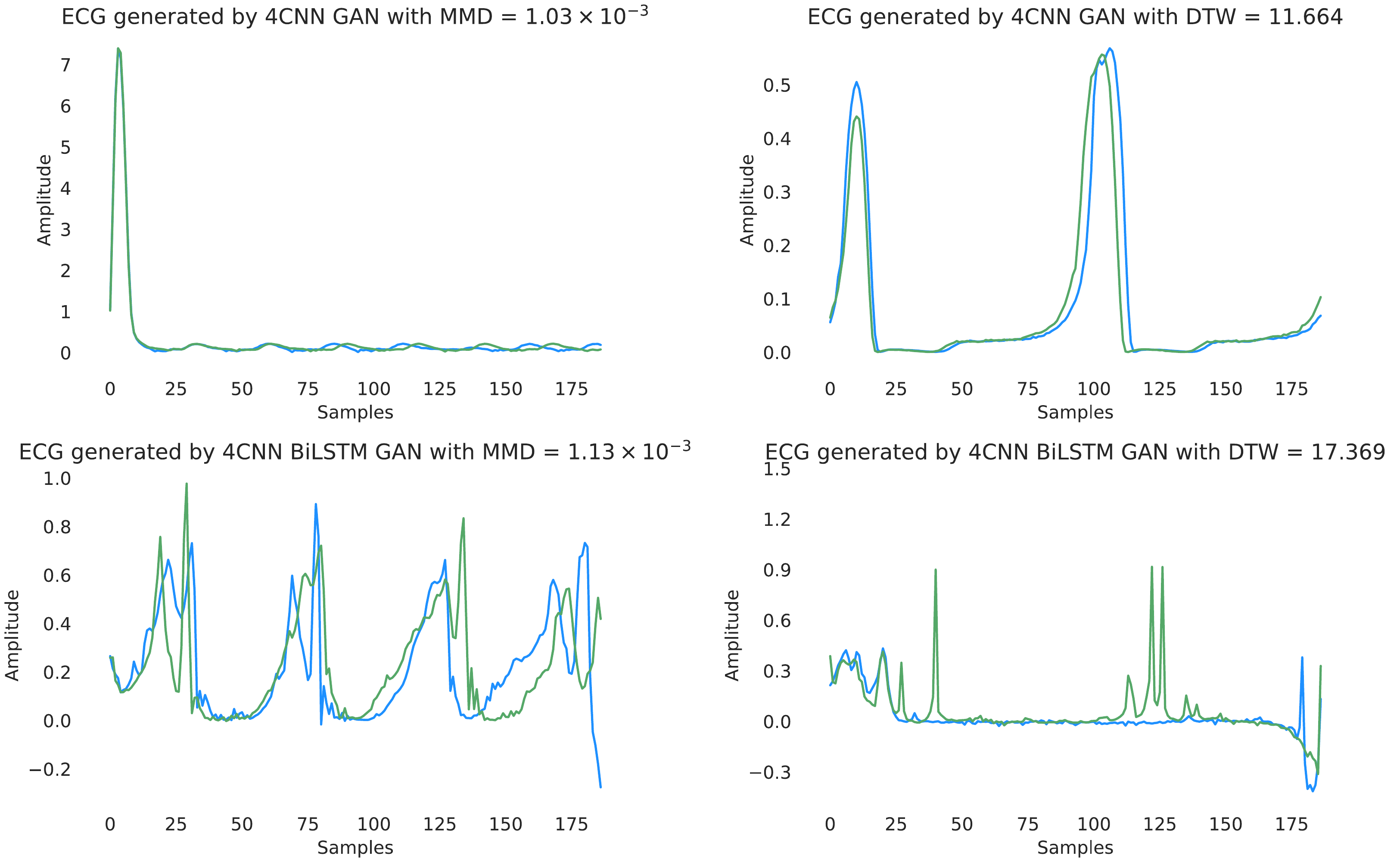}}
\caption{\label{fig:ECGResults}Examples of ECG waves corresponding to the MMD and DTW reported in Table \ref{tab:ECGMetrics}.}
\end{figure}

As expected, ECG proved much more difficult to synthesise than sinusoidal data. Both the 4CNN GAN and the 4CNN BiLSTM GAN were highly unstable during training, with the majority of experiments including minibatch discrimination layers failing and the errors exploding. The best results for MMD and DTW are reported in Table \ref{tab:ECGMetrics} for both the 4CNN GAN and the 4CNN BiLSTM GAN. The corresponding ECG waves are displayed in Figure \ref{fig:ECGResults}. Note that only results where MMD $>0.001$ are included here to avoid examining results where MMD failed. In contrast to the sine data, DTW proved a more effective evaluation metric than MMD for ECG data. As mentioned previously, DTW is more sensitive to the relative scale between the synthetic and test data in comparison to MMD. For example, while data generated using the 4CNN GAN produced the best MMD score of all experiments, it is clear from Figure \ref{fig:ECGResults} that the scale is incorrect in comparison to the training and test data. Thus, DTW proves more robust in the situation where training fails and the range of the synthetic data increases unbounded. The output of the all GANs were visually evaluated to determine if better results were produced during training than identified by our evaluation metrics. The 4CNN BiLSTM GAN was used to generate the data in Figure \ref{fig:RealvFakeECG} which is similar in form to real ECG.

\begin{table}[!ht]
\begin{center}
\begin{tabular}{|c| c| c|} 
 \hline
 GAN &  MMD (mb)& DTW  (mb)\\
 \hline\hline
 4CNN GAN & $\mathbf{1.03\times 10^{-3} (5)}$ & \textbf{11.664 (0)}\\ 
4CNN BiLSTM GAN & $1.13\times 10^{-3} (0)$ & 17.369 (0) \\
 \hline
\end{tabular}
\caption{\label{tab:ECGMetrics}The best MMD and DTW scores obtained for ECG data generated by each GAN is listed. The number of outputs in the minibatch discrimination layer is given in brackets. The best MMD and DTW overall are in bold.}
\end{center}
\end{table}

\subsubsection{Minibatch Discrimination} 
\begin{figure}[!htb]
    \center{\includegraphics[width=0.85\columnwidth]
    {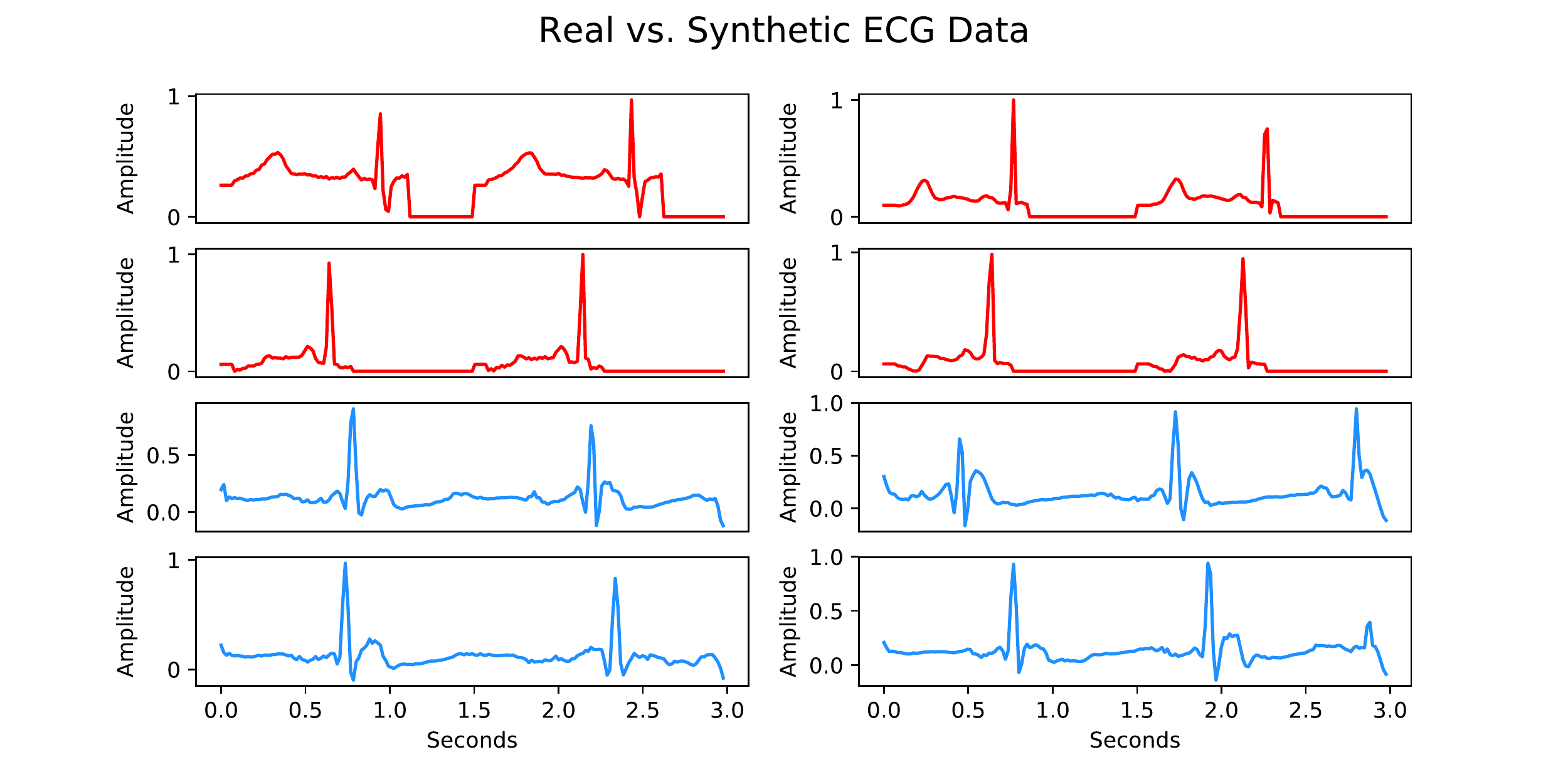}}
\caption{\label{fig:RealvFakeECG}Examples of real (red) and synthetic (blue) ECG waves. }
\end{figure}

It was difficult to study the effect of the inclusion of a minibatch discrimination layer on the synthetic ECG produced as the majority of experiments failed to converge during training. This increased instability could be caused by the inclusion of four convolution-pooling layers in the discriminator. While it was not possible to study GANs which contained a minibatch discrimination layer in the discriminator, it is worth noting that mode collapse did not occur for the 4CNN BiLSTM GAN when no minibatch discrimination layer was included. The diverse results produced by this GAN can be seen in Figure \ref{fig:ECGResults} and Figure \ref{fig:RealvFakeECG}. It is possible that the complexity of the ECG data prevents the generator from learning a select number of features which fool the discriminator, thus preventing the occurrence of mode collapse. Further work is required to stabilise these GANs to enable a complete study of minibatch discrimination.

\subsubsection{Presence Disclosure}
The signals in Figure \ref{fig:RealvFakeECG} produced by the 4CNN BiLSTM GAN were tested to determine if the GAN was successful in protecting the privacy of the underlying training data. 
The results show that the number of training records identified is low, with approximately 0\% correctly identified for $\epsilon<0.3$ $\times$ \textit{mean distance}. However, as the thresholds increase above this boundary, the number of records correctly identified as training records increases independently of the sample size $r$. As before, precision is approximately 50\% for all $\epsilon$ and $r$, but fluctuates when $\epsilon=0.3$ $\times$ \textit{mean distance}. Overall, this is a promising result, showing that for complex data, GANs can produce synthetic data which protects the privacy of the original data.

\begin{figure}[!htb]
\centering
{\includegraphics[width=0.75\columnwidth]{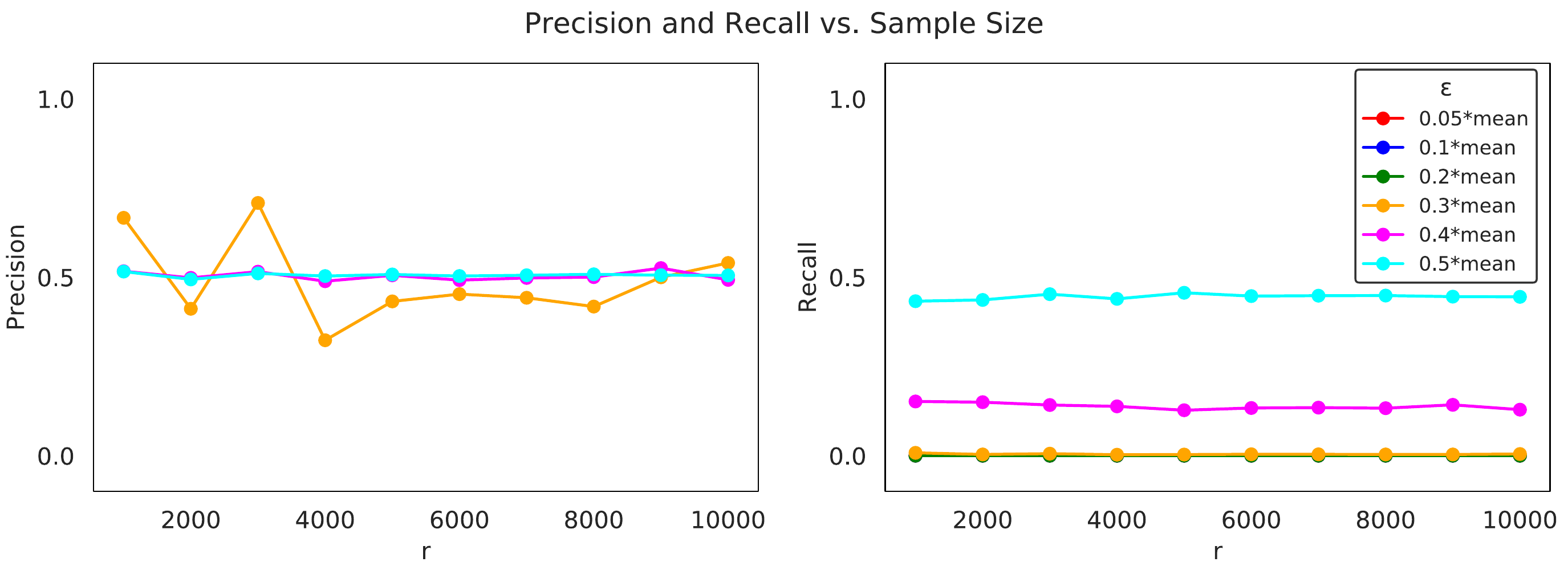}}
\caption{The precision and recall scores for the presence disclosure test using ECG data show that GANs can protect the privacy of the training data, especially for low $\epsilon$.}
\label{fig4:ECGPresence}
\end{figure}

\section{Future Work}
As discussed previously, the training phase of GANs is highly unstable and more research is required to stabilise this process. The implementation of alternative loss functions such as the Wasserstein GAN, or the inclusion of normalisation layers should be investigated. By stabilising the training process, more rigorous methods could be used to determine the optimum parameters for generating time series data. It is clear from our preliminary research that GANs are a promising method for generating synthetic time series. However, there is still a need to implement metrics such as DTW for privacy evaluation to further investigate suitable methods to assess the quality and privacy of the data produced. Possible alternatives include the classifier two-sample test and comparison of Fourier transforms of the series. Finally, further tests are required to to assess the ability of GANs to reduce privacy risks. Differential private GANs should also be examined for this purpose. 

\section{Conclusion}
To address growing privacy concerns related to the sharing of sensitive medical data, an alternative method for generating synthetic time series suitable for use in clinical training or further research was investigated. We developed a set of GANs capable of generating both synthetic sine waves and ECG waves. A GAN consisting of two bidirectional LSTMs in the generator and convolution-pooling layers in the discriminator generated high-quality data in both cases. Furthermore, we demonstrated that the inclusion of a minibatch discrimination layer in the discriminator can prevent mode collapse in training. To contribute to the continuing search for suitable evaluation metrics for GANs, both MMD and DTW were studied as part of this work. It was shown that MMD favours GANs which generate a diverse range of outputs while DTW is more robust against training instability due to its sensitivity to the relative amplitude between the real and synthetic data. Finally, the GANs presented in this paper demonstrate promising results for the protection of data privacy for membership inference attacks. However, further research is required to stabilise GANs and to ensure that they reduce the privacy risk associated with sharing sensitive medical time series data.

\section*{Acknowledgement}
We gratefully acknowledge the support of NVIDIA Corporation with the donation of the Titan Xp used for this research.

\bibliographystyle{IEEEtran}  
\bibliography{references}  




\end{document}